\documentclass{article}

\usepackage{amsmath, amssymb}
\usepackage{graphicx}
\usepackage{float}
\usepackage{hyperref}
\hypersetup{colorlinks=true,citecolor=blue,urlcolor=blue,linkcolor=blue}
\usepackage{authblk}
\usepackage{subcaption}
\usepackage[a4paper, total={6in, 9in}]{geometry}
\usepackage{multirow}

\DeclareMathOperator*{\argmax}{arg\,max}

\renewenvironment{abstract}{%
	\begin{center}
		\begin{minipage}{0.75\textwidth}
			\footnotesize
			\textbf{Abstract: }
		}{%
		\end{minipage}
	\end{center}
	\vspace{5mm}
}

\title{Using Graph Neural Networks for the segmentation of overlapping objects in high granularity calorimeters}

\author[1]{Matthieu Melennec}
\author[1]{Fr\'ed\'eric Magniette}

\affil[1]{Ecole Polytechnique, IN2P3-CNRS, Laboratoire Leprince-Ringuet, F-91120 Palaiseau, France}

\date{}

\begin{document}
	
	\maketitle
	
	\begin{abstract}

		
		High-granularity calorimeters at the High-Luminosity LHC require novel algorithms to resolve overlapping particle showers. We present an optimised Graph Neural Network (GNN) segmentation block that predicts node-level energy fractions to disentangle overlapping two-photon showers. By accelerating graph construction and convolution operations, our pipeline achieves competitive separation efficiency with significantly reduced algorithmic complexity.
	\end{abstract}

	\section{Introduction}
	
	The High-Luminosity LHC (HL-LHC) upgrade \cite{aberle_20} will induce severe pileup with nearly 200 simultaneous collisions per bunch crossing, causing spatial overlapping of energy deposits in conventional calorimeters. To overcome this limitation and disentangle individual particle showers, next-generation experiments are deploying highly granular detectors, such as the CMS High Granularity Calorimeter (HGCAL) \cite{pasztor_23, cms_17, atlas_17}.
	
	To fully exploit the unprecedented level of precision offered by this high granularity image of the shower development, new algorithms must be developed \cite{cms_17}. While deep learning based reconstruction algorithms have shown their success on the conventional calorimeter geometries \cite{simkina_24, maidannyk_26}, they can be hard to generalise to high dimensional outputs with variable granularity, such as the future HGCAL. Graph Neural Networks (GNN) \cite{gilmer_17} are promising candidates for particle reconstruction in these complex environments \cite{qasim_19, bhattacharya_23}. However, they rely on algorithms that are prohibitively expensive to use. For instance, graph construction can usually only be done in a quadratic number of operations in the number of nodes \cite{qasim_19, franceschi_19}, and is often used repeatedly within the same pipeline.
	
	We present how GNN based pipelines can be optimised to exploit the granularity of new generation detectors to segment overlapping objects in granular detectors. In particular, we introduce SOOP (Segmentation of Overlapping Objects Pipeline), an optimised GNN pipeline that reduces the algorithmic complexity of standard GNNs that dynamically update adjacencies. We evaluate its performance on overlapping photon showers, as can be observed from the decay of neutral pions \cite{pdg_24}.
	
	\section{Event Simulation}
	
\begin{figure}
		\centering
		\begin{subfigure}{0.45\textwidth}
			\centering
			\includegraphics[width=\textwidth]{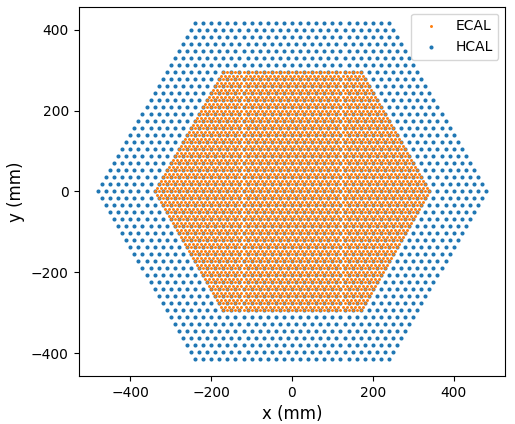}
		\end{subfigure}
		\begin{subfigure}{0.51\textwidth}
			\centering
			\includegraphics[width=\textwidth]{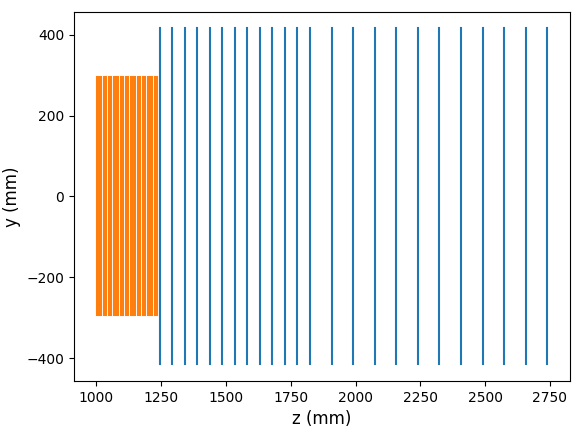}
		\end{subfigure}
		\caption{The D2 detector geometry. We observe the two sections of the detector, with the more granular ECAL (orange), followed by the HCAL (blue). We observe in particular that the HCAL has two subsections with different longitudinal granularity.}\label{sec:sim-fig:d2_geom}
	\end{figure}

	The data used in this work is obtained from the D2 dataset \cite{becheva_24}. It is a set of simulated events in a sampling calorimeter with an HGCAL-inspired structure. This dataset is based on \textsc{g\'eant}4 \cite{geant4_16} to simulate the passage of particles through this detector. It is comprised of 26 electromagnetic calorimeter (ECAL) layers, followed by a hadronic calorimeter (HCAL) comprised of 24 layers. Figure \ref{sec:sim-fig:d2_geom} shows the geometry of the detector.
	
	The D2 dataset contains samples of single photons, shot 1 m away from the first ECAL layer. Their energies range between 10 and 100 GeV, and their directions form an angle  between $0^\circ$ and $15^\circ$ with the longitudinal axis of the detector. The resulting events provide a list of hits, containing the hit sensor's position and read energy. Overlapped photon pair events are constructed by overlapping individual photon events, such that shared cells take the sum of deposited energies. To focus on non-trivial overlapping showers, events are restricted to angular separations $\varphi \le 4.5^\circ$. To ensure that showers are separated enough for the model to be able to segment them, we impose a lower bound $\varphi\ge 1^\circ$.
	
	\begin{figure}
		\begin{minipage}{0.48\textwidth}
			\centering
			\includegraphics[width = \textwidth]{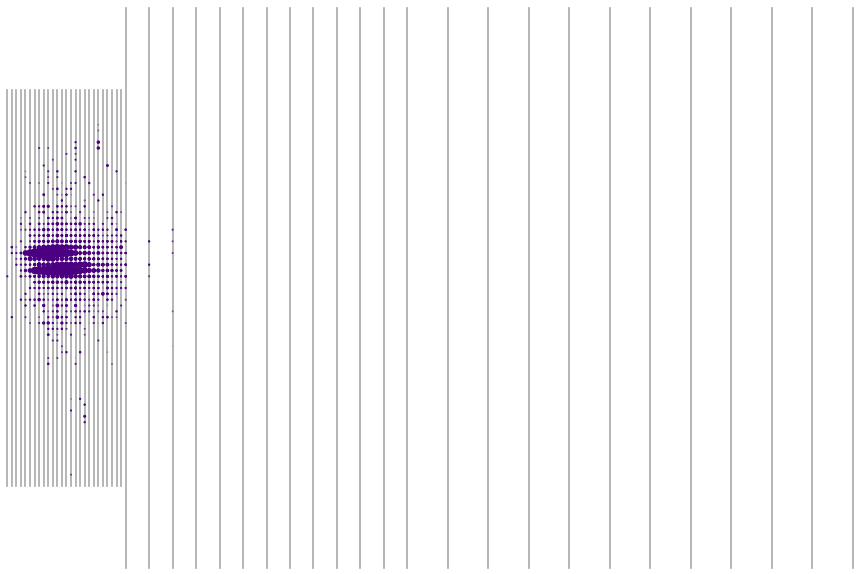}
			\caption{Typical point cloud obtained from a well resolved photon pair event. One recognises the two shower structures from either photon interactions with the detector.}\label{sec:sim-fig:evt}
		\end{minipage}
		\hfill
		\begin{minipage}{0.45\textwidth}
			\centering
			\includegraphics[width=\textwidth]{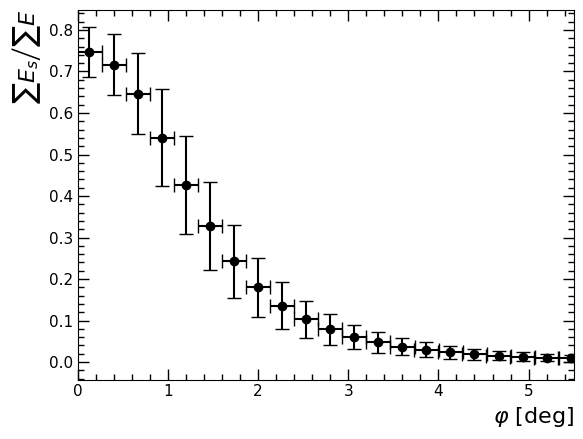}
			\caption{Ratios of the shared deposited energy and the total deposited energy by photons in the D2 detector, as a function of separation angle $\varphi$.}\label{sec:sim-fig:shared_energy}
		\end{minipage}
	\end{figure}
	
	\section{Model Structure}
	
	The objective of this pipeline is to regress, for each hit, the fraction $f_{A,B}$ of its energy that belonged to shower $A$ or $B$, triggered by either photons $\gamma_A$ and $\gamma_B$ of the event, respectively. The individual showers are then reconstructed from these fractions by weighting the hit energies by $f_A$ and $f_B$ respectively.
	
	\subsection{Graph Construction}
	For each event, we first construct a graph from the overlapping shower data.. Node features consist of 3D Euclidean coordinates and cell energy $v = (x, y, z, E) \in \mathbb{R}^3 \times \mathbb{R}^+$. Graphs are constructed using the PT-KNN algorithm \cite{melennec_25} ($k=4$) with edge features $e_{vw} \in \mathbb{R}^+$ defined by cell separation. The PT-KNN algorithm speeds up the graph construction process by using pre-computed proximity tables where detector sensors are pre-ordered in increasing distance. This design avoids full pairwise distance calculations during event processing, greatly reducing execution time.
	
	\subsection{Node-Level Energy Fraction Regression}\label{sec:pipeline-subsec:node_level_regression}
	
	A direct formulation of the task would assign each node to shower $A$ or shower $B$. The choice of labels $A$ or $B$ is inherently symmetric, and thus completely arbitrary. This makes a direct $A/B$ classification target ill-defined in the absence of an ordering convention. This global permutation ambiguity means that there is no physically meaningful ordering that uniquely identifies one shower as $A$ and the other as $B$. We therefore formulate the problem locally. For each node, the network predicts (i) the fraction of energy deposited by the locally dominant shower and (ii) the estimated detector entry point of that shower. The predicted entry points are then clustered into two groups, which provide a consistent event-level assignment of nodes to the two incident photons. The corresponding energy fractions are finally used to reconstruct the two showers. For each node $v\in V$, we estimate the biggest fraction, i.e. the fraction $f_v^{(X)}\in [0.5, 1]$ of energy from the particle contributing the largest energy fraction to that cell, and, on the other side, the event level assignment, through the position $\mathbf{u}_v^{(X)}\in \mathbb{R}^2$ at which the particle that deposited the biggest fraction of energy entered the detector. Here, $X$ denotes the label $A$ or $B$ corresponding to the shower that deposited the biggest amount of energy in that cell, i.e. $f_v^{(X)} \geq f_v^{\left(\bar{X}\right)}$. From the predicted position of entry, we cluster the nodes in two sets, one for each shower, and assign the energy of each hit using the predicted fraction. Assuming exact predictions for $f_v^{(X)}$ and $\mathbf{u}_v^{(X)}$, this information is sufficient to reconstruct both showers: the dominant fraction is assigned to the shower associated with the predicted entry point, while the complementary fraction is assigned to the other shower:
	\begin{equation*}
		V^{(A)} = \left\{ \left. E_v f_v^{(X)} \right\vert X=A \right\} \bigcup \left\{ \left. E_v \left(1-f_v^{(X)}\right) \right\vert X=B \right\},
	\end{equation*}
	and vice-versa for $B$. This dual regression objective is essential to successfully segment the overlapping objects, as it structurally provides a symmetric fraction estimation, respecting the symmetries of the input structures, while the position of entry estimation localises the nodes, enhancing the separation of individual nodes. This approach thus fully clears the global ambiguity from the permutation of both showers.
	
	To obtain the estimations for $f_v^{(X)}, \mathbf{u}_v^{(X)}$, the graph is first fed into consecutive graph convolution layers. The convolution operation is a message passing convolution \cite{gilmer_17}, implemented as
	\begin{equation}
		x_v^{(l+1)} = \underset{w\in \tilde{\mathcal{N}}(v)}{\square}
		\phi_\Theta \left( \text{concat}\left[x_v^{(l)}, x_w^{(l)}, e_{vw} \right] \right)
	\end{equation}
	The $\phi_\Theta$ is an MLP with trainable weights $\Theta$, and a Leaky-ReLU activation. We use feature-wise mean aggregation, following the configuration adopted in Ref. \cite{goodfellow_16} for regression tasks. Because this task is a node-level regression, we do not implement any pooling layer in the GNN block. Since the graph size is preserved while the feature dimensionality increases, we use concatenation skip connections \cite{huang_17} to facilitate information and gradient propagation. The output features for each node after the convolution layers are passed through an MLP, that will output $p_v\in [0.5, 1]$, the biggest predicted fraction from either shower, and a position $\tilde{\mathbf{u}}_v\in \mathbb{R}^2$. The distribution of predicted entry points $\tilde{U} = \tilde{\mathbf{u}}_{v\in V}$ is then used to estimate the two incident-particle entry positions $\mathbf{u}^{(\chi)}, \chi \in \{ A, B \}$.
	\begin{figure}
		\centering
		\includegraphics[width=0.65\textwidth]{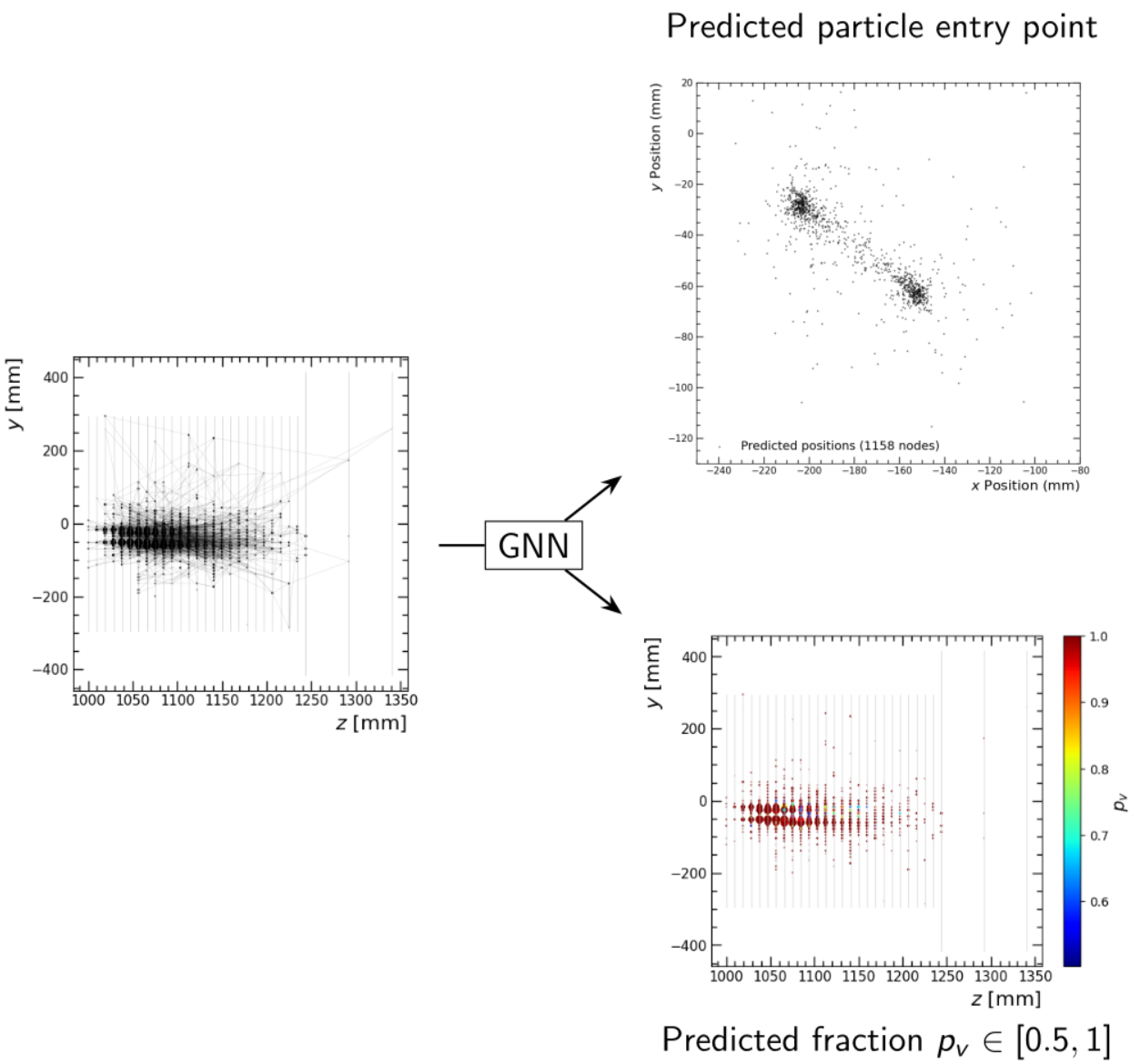}
		\caption{The GNN part of the SOOP pipeline predicts, for each node $v$, the position of entry in the detector of the particle responsible for the biggest energy deposition in node $v$, and the fraction of the energy $E_v$ from that particle.}
	\end{figure}
	
	To obtain this proxy pair, we fit $\mathbf{u}^{(\chi)}$ by training a Gaussian Mixture Model (GMM) \cite{pearson_1894, dempster_77} with 2 Gaussians:
	\begin{equation*}
		\tilde{U} \sim \sum_\chi w^{(\chi)} \mathcal{N}\left( \mathbf{u}^{(\chi)}, \Sigma^{(\chi)} \right),
	\end{equation*}
	and use the averages as proxies for $\mathbf{u}^{(X)}$. Because of the possible imbalance in node count and sparsity of the data, an unconstrained GMM does not reliably identify both shower centres. We thus	first remove outliers with a DBSCAN, and enforce a tied covariance matrix $\Sigma = \Sigma^{(\alpha)} = \Sigma^{(\beta)}$.
	\begin{figure}
		\centering
		\includegraphics[width = 0.45\textwidth]{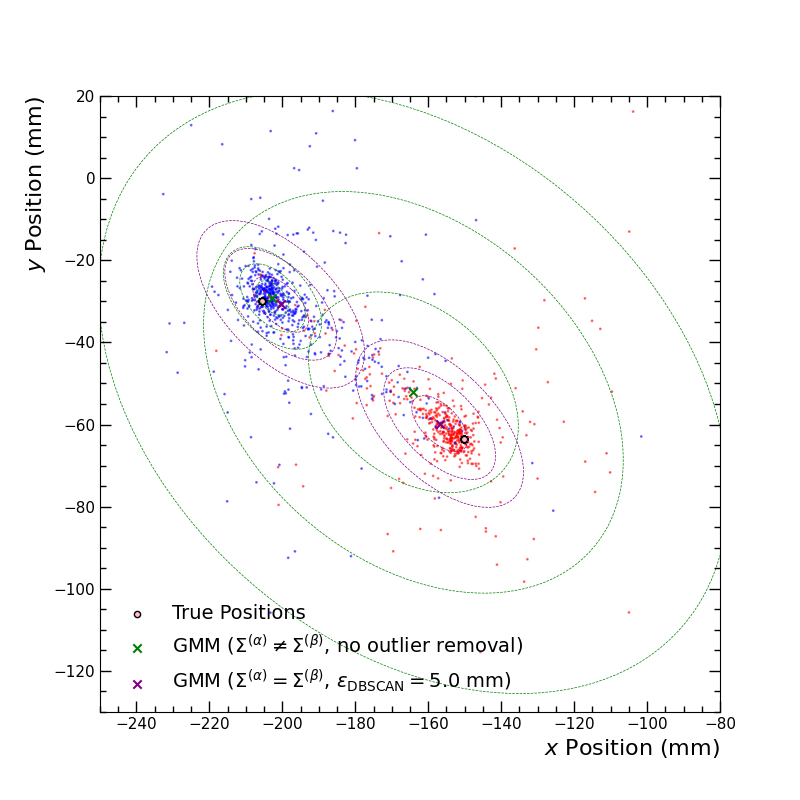}
		\caption{A simple GMM often fails to account for the imbalance in node counts and sparsity of the data. Adding a DBSCAN for outlier removal, and imposing a tied covariance matrix ensures both clusters are fitted.}\label{sec:pipeline-fig:gmm}
	\end{figure}
	
	Once the GMM model is fitted to the data, we assign each node to either particle $\alpha$ or $\beta$ from the most probable source of the Gaussian mixture,
	\begin{equation*}
		\argmax_{\chi} w^{(\chi)} \mathbb{P}\left[ \tilde{\mathbf{u}}_v \left\vert \mathbf{u}^{(\chi)}, \Sigma \right.\right].
	\end{equation*}
	Each shower is then reconstructed as
	\begin{equation}
		V^{(\chi)} = \left\{ \left(\mathbf{x}_v, E_vp_v^{(\chi)}\right) \right\}, \qquad
		p_v^{(\chi)} = \begin{cases}
			p_v & \text{if }\tilde{\mathbf{u}}_v \sim \mathbf{u}^{(\chi)}, \\
			1-p_v & \text{if } \tilde{\mathbf{u}}_v \sim \mathbf{u}^{(\bar{\chi})}.
		\end{cases}
	\end{equation}
	
	In summary, the complete pipeline operates in several distinct stages. First, the GNN predicts both a dominant energy fraction and a corresponding detector entry position for every individual hit. Next, these predicted positions are clustered using a DBSCAN outlier reduction and a GMM to determine an clustered position of entry estimation. Finally, this event-level assignment is combined with the predicted fractions to reconstruct the isolated energy deposits of the original single showers.
	
	This clustering is the principal bottleneck of the SOOP pipeline. It imposes that a Gaussian Mixture Model be fitted to the GNN's output at inference. It is also not easily scalable, which is prohibitive in high pileup environments, like in the HGCAL detector in the HL-LHC era. The main difficulty in using pre-trained models is that the clustering space being the spatial coordinates of entry of the particles, the latent space features vary drastically from one sample to another, not allowing any clustering model to be pre-trained. Using a clustering routine that is more scalable, and ideally differentiable, is thus a natural approach to improve this model.
	
	\section{Training and Performance}
	
	To avoid performing GMM clustering at every training iteration, we train the model on the convolution layers output, i.e. particle entry position and predicted fraction. To this end, we define a loss function comprising two terms, one for each regression objective
	\begin{gather}
		\mathcal{L} = \lambda_\text{pos} \mathcal{L}_\text{pos} + \lambda_\text{freq} \left( \mathcal{L}_\text{frac}^{(A)} + \mathcal{L}_\text{frac}^{(B)} \right), \\
		\mathcal{L}_\text{pos} = \frac{1}{\vert V \vert} \sum_{v\in V} \left\Vert \tilde{\mathbf{u}}_v - \mathbf{u}^{(X_v)} \right\Vert^2, \qquad \mathcal{L}_\text{frac}^{(X)} = \frac{\sum_{v\in V} \sqrt{E_v f_v^{(X)}} \left( f_v^{(X)} - p_v^{(X)} \right)^2}{\sum_{v\in V} \sqrt{E_v f_v^{(X)}}},
	\end{gather}
	where $X_v\in \{A, B\}$ is the label of the principal incident particle of $v\in V$.
	The loss term on the fraction prediction is adapted from \cite{qasim_19}, which works on background segmentation for pions. An energy normalisation is performed to ensure normalised loss values across events across variable sizes. The energy weighting of the fraction prediction square error reduces the contribution of low-energy hits, which are not as relevant as the shower core, and may induce significant noise.
	
	The loss weights were chosen so that all loss terms would have the same order of magnitude, giving $\lambda_\text{pos} = 10^{-5}$ and $\lambda_\text{frac}=1$. The model was trained with $6\times 10^5$ parameters, over 10M graphs. With a learning rate of $10^{-3}$ and Adam optimiser, it converged in $\sim 100$ epochs.
	
\begin{figure}
		\begin{minipage}{0.46\textwidth}
			\centering
			\includegraphics[width=\textwidth]{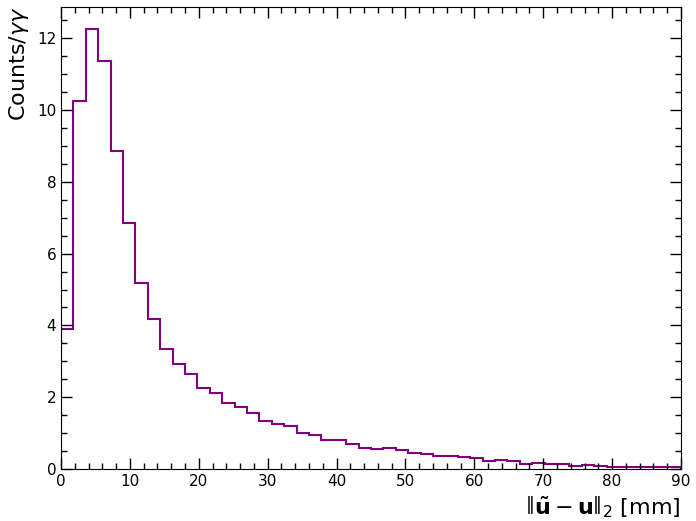}
			\caption{Distance distribution between the true and reconstructed position of entry of the incident $\gamma$s.}\label{sec:perf-fig:reco_dist}
		\end{minipage}
		\hfill
		\begin{minipage}{0.48\textwidth}
			\centering
			\includegraphics[width=\textwidth]{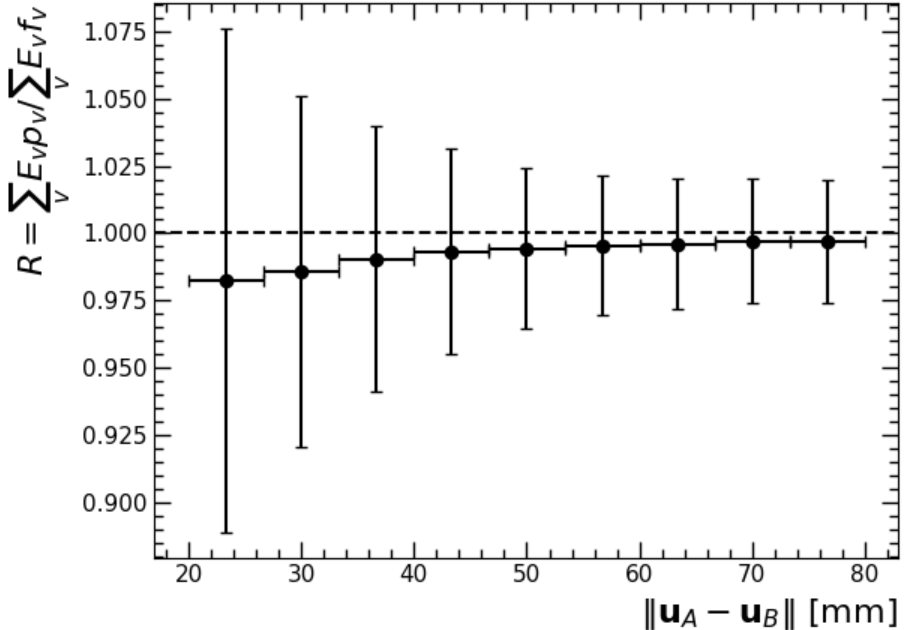}
			\caption{SOOP response as a function of distance between the entry points of incident $\gamma$s.} \label{sec:perf-fig:resp}
		\end{minipage}
	\end{figure}

	We first examine the distribution of distances between the means obtained by the GMM fitting and the actual entry-points of incident particles. It is shown in figure \ref{sec:perf-fig:reco_dist}. 
	This figure shows that while the reconstructed entry positions remain concentrated near the true positions, we observe a bias, typically towards the halfway mark between the position of entry of the incident photons (as can be observed in figure \ref{sec:pipeline-fig:gmm}). Note that this effect could be mitigated by performing the clustering during the forward passes of the training, as it would permit us to apply the position loss on the cluster centres, allowing the model to consequently bias single node predictions further out.
	
	To evaluate the performance of the  model, we use the response function derived from \cite{qasim_19},
	\begin{equation}
		R = \frac{R^{(A)} + R^{(B)}}{2}, \qquad R^{(X\in \{A, B\})} = \frac{\sum_{v\in V} E_vp_v^{(X)}}{\sum_{v\in V} E_vf_v^{(X)}},
	\end{equation}
	This response function evaluates the reconstruction capacities of the global energy of each shower. Observe how it does not take into account the exact distribution of node reconstructed energies, as we sum  over the nominator and denominator, instead of using the average response per node. This choice was made because the end goal of this segmentation is not to recover the exact shower structure, but rather to reconstruct a point cloud such that meaningful information can be retrieved from it. As such, with this objective, it isn't as important to reconstruct two small nodes in the periphery of the shower as to recover one centroid with the total energy from both nodes. Getting the total deposited energy is however crucial, since the incident energy $E_\gamma$ scales with $\sum_{v\in V} E_v$ \cite{wigmans_00}.
	
	We look at the distribution of responses, binned in distance between particle entry-points. Further entry points mean that the two showers should be better resolved, and as such we observe an improvement in the response. We fit the binned distributions of responses by Gaussian functions. This result is shown in figure \ref{sec:perf-fig:resp}, where we look at the average response and standard deviations of the Gaussian fits.

	We further observe that the standard deviations of responses improve significantly. This is due to the decreased amount of shared hits as distance increases (see figure \ref{sec:sim-fig:shared_energy}), and as such, a lot less nodes have fractions $f^{(A)}\approx f^{(B)} \approx 0.5$, which are the hardest values to regress. Furthermore, the outermost nodes have a much smaller energy deposition, and as such take a lot less part in the response value.
	
	Observe that static geometry-aware message passing achieves competitive segmentation performance without dynamic adjacency updates \cite{qasim_19}. While these are applied to noise reduction in granular hadronic calorimeters, this shows that our method, despite using static edges implementing locality in the detector geometry, recovers the same abstraction power as computationally heavy object condensation methods.
	
	On average, the response $\langle R\rangle$ is smaller than 1, indicating a residual bias in the per-shower energy reconstruction. Once again, if the clustering operation were included during training, we could use a loss term to penalise under-estimations of the energy, allowing the loss function to directly penalise this bias.
	
	\section{Conclusion}
	
	We presented the SOOP pipeline, an optimised Graph Neural Network approach designed to segment overlapping shower pairs in high-granularity calorimeters such as the CMS HGCAL. SOOP replaces computationally heavy graph construction with pre-computed proximity tables and using localised message-passing convolutions. This adjacency evaluation reduces to sub-quadratic complexity while retaining sufficient geometric information to disentangle overlapping showers. The proposed node-level energy fraction regression shows that GNNs can perform detailed spatial disentanglement and recover incident particle entry points beyond simple background rejection. The achieved separation performance is comparable with state-of-the-art dynamic methods while reducing algorithmic complexity.
	
	Despite these results, two structural limitations must be addressed. First, SOOP relies on a non-differentiable Gaussian Mixture Model for post-hoc clustering. This prevents a full optimisation with the total shower reconstruction loss. As such, it needs future work on pre-trainable, standardised latent spaces. Secondly, SOOP is a node-level regression pipeline, which does not implement any graph pooling. This causes an attenuation of messages from further nodes, which limits a multi-scale representation, making the information flow less efficient. Using multi-scale architectures, such as graph encoder-decoders \cite{li_21}, should mitigate this attenuation and further improve segmentation performance.
	
	\section*{Acknowledgments}
	
	The authors are thankful to the ANR for the funding of the OGCID project, of which this study is part of, through funding ANR-21-CE31-0030.
	
	\bibliographystyle{ieeetr}
	\bibliography{biblio}

@book{aberle_20,
	author = {O. Aberle and others},
	title = {High-Luminosity Large Hadron Collider (HL-LHC): Technical design report},
	publisher = {CERN},
	series = {CERN Yellow Reports: Monographs},
	volume = {10},
	year = {2020},
	doi = {10.23731/CYRM-2020-0010}
}

@article{pasztor_23,
	author = {G. {Pásztor on behalf of the CMS Collaboration}},
	title = {The Phase-2 Upgrade of the CMS Detector},
	journal = {Proceedings of Science},
	volume = {LHCP2022},
	year = {2023},
	pages = {045},
	doi = {10.22323/1.422.0045}
}

@techreport{atlas_17,
	author = {{The ATLAS Collaboration}},
	title = {Technical Design Report for the Phase-II Upgrade of the ATLAS TDAQ System},
	institution = {CERN},
	year = {2017},
	doi = {10.17181/CERN.2LBB.4IAL}
}

@techreport{cms_17,
	author = {{The CMS Collaboration}},
	title = {The Phase-2 Upgrade of the CMS Endcap Calorimeter},
	institution = {CERN},
	year = {2017},
	doi = {10.17181/CERN.IV8M.1JY2}
}

@inproceedings{gilmer_17,
	author = {J. Gilmer and others},
	title = {Neural message passing for Quantum chemistry},
	booktitle = {Proceedings of the 34th International Conference on Machine Learning},
	volume = {70},
	year = {2017},
	pages = {1263-1272},
	doi = {10.48550/arXiv.1704.01212}
}

@article{qasim_19,
	author = {S. R. Qasim and others},
	title = {Learning representations of irregular particle-detector geometry with distance-weighted graph networks},
	journal={The European Physical Journal C},
	volume = {79},
	year = {2019},
	pages = {608},
	doi = {10.1140/epjc/s10052-019-7113-9}
}

@article{bhattacharya_23,
	author = {S. Bhattacharya and others},
	title = {GNN-based end-to-end reconstruction in the CMS Phase 2 High-Granularity Calorimeter},
	journal = {Journal of Physics: Conference Series},
	volume = {2438},
	year = {2023},
	pages = {012090},
	doi = {10.1088/1742-6596/2438/1/012090}
}

@inproceedings{franceschi_19,
	author = {L. Franceschi and others},
	title = {Learning Discrete Structures for Graph Neural Networks},
	booktitle = {Proceedings of the 36th International Conference on Machine Learning},
	volume = {97},
	year = {2019},
	pages = {1972-1982},
	doi = {10.48550/arxiv.1903.11960}
}

@article{pdg_24,
	author = {Particle Data Group},
	title = {Review of particle physics},
	journal = {Physical Review D},
	volume = {110},
	year = {2024},
	pages = {030001},
	doi = {10.1103/PhysRevD.110.030001}
}

@misc{becheva_24,
	author = {E. Becheva and others},
	title =  {High Granularity Calorimetry D2 Dataset},
	year = {2024},
	url = {https://zenodo.org/records/14260279}
}

@article{geant4_16,
	author = {J. Allison and others},
	title = {Recent Developments in \textsc{géant}4},
	journal = {Nuclear Instruments and Methods in Physics Research Section A: Accelerators, Spectrometers, Detectors and Associated Equipment},
	volume = {835},
	year = {2016},
	pages = {186-225},
	doi = {10.1016/j.nima.2016.06.125}
}

@article{melennec_25,
	author = {M. Melennec and S. Ghosh and F. Magniette},
	title = {Optimised Graph Convolution for Calorimetry Event Classification},
	journal = {European Physics Journal Web of Conferences},
	volume = {337},
	year = {2025},
	pages = {01024},
	doi = {10.1051/epjconf/202533701024}
}

@book{goodfellow_16,
	author = {I. Goodfellow and Y. Bengio and A. Courville},
	title = {Deep Learning},
	publisher = {MIT Press},
	year = {2016},
	url = {http://www.deeplearningbook.org}
}

@inproceedings{huang_17,
	author = {G. Huang and others},
	title = {Densely Connected Convolutional Networks}, 
	booktitle = {IEEE Conference on Computer Vision and Pattern Recognition},
	year = {2017},
	pages = {2261-2269},
	doi = {10.1109/CVPR.2017.243}
}

@article{pearson_1894,
	author = {K. Pearson},
	title = {Contributions to the mathematical theory of evolution},
	journal = {Philosophical Transactions of the Royal Society of London. (A.)},
	volume = {185},
	year = {1894},
	pages = {71-110},
	doi = {10.1098/rsta.1894.0003}
}

@article{dempster_77,
	author = {A. P. Dempster and N. M. Laird and D. B. Rubin},
	title = {Maximum Likelihood from Incomplete Data Via the EM Algorithm},
	journal = {Journal of the Royal Statistical Society B},
	volume = {39},
	year = {1977},
	pages = {1-22},
	doi = {10.1111/j.2517-6161.1977.tb01600.x}
}

@book{wigmans_00,
	author = {R. Wigmans},
	title = {Calorimetry: Energy Measurement in Particle Physics},
	publisher = {Clarendon Press},
	series = {International series of monographs on physics},
	year = {2000},
	doi = {10.1093/oso/9780198786351.001.0001}
}

@inproceedings{li_21,
	author={J. Li and others},
	title={Deconvolutional Networks on Graph Data},
	booktitle={Advances in Neural Information Processing Systems},
	year={2021},
	doi = {10.48550/arXiv.2110.15528}
}

@article{simkina_24,
	author = {P. Simkina and others},
	title = {Reconstruction of electromagnetic showers in calorimeters using Deep Learning},
	journal = {European Physics Journal C},
	volume = {84},
	pages = {639},
	year = {2024}
}

@article{maidannyk_26,
	author = {Y. Maidannyk and others},
	title = {Reconstruction of overlapping electromagnetic showers in calorimeters using Transformers},
	journal = {European Physics Journal C},
	volume = {86},
	pages = {873},
	year = {2026}
}

\end{document}